\documentclass[10pt,conference,a4paper]{IEEEtran}

\usepackage[utf8]{inputenc}

\usepackage{blindtext, graphicx}
\usepackage[utf8]{inputenc}
\usepackage[english]{babel}
\usepackage{csquotes}
\usepackage{wrapfig}
\usepackage{float}
\usepackage{verbatim}
\usepackage{subcaption}
\usepackage[font=footnotesize,labelfont=bf]{caption}
\usepackage{import}
\usepackage{dirtytalk}
\usepackage{subfiles}
\usepackage{array}
\usepackage{multirow}

\usepackage{lscape}
\usepackage{algorithm}
\usepackage{algorithmic}
\usepackage{rotating}
\usepackage[table,xcdraw]{xcolor}

\usepackage{longtable} 
\usepackage{array} 
\usepackage{caption} 
\usepackage{graphicx} 
\usepackage{xcolor, colortbl} 
\usepackage{booktabs} 
\usepackage{tabularx} 
\usepackage{amsmath}

\usepackage{times}

\DeclareGraphicsExtensions{.png,.eps,.ps,.pdf}

\usepackage{url}
\usepackage{hyperref}

\usepackage[most]{tcolorbox}

\usepackage{multirow}
\usepackage{tabularx}
\usepackage{makecell}
\usepackage{svg}

\usepackage{pifont}
\definecolor{ao}{rgb}{0.0, 0.5, 0.0}
\definecolor{amber}{rgb}{1.0, 0.49, 0.0}

\let\labelindent\relax
\usepackage{enumitem}

\usepackage{lscape} 

\graphicspath{{Figures/}}

\usepackage{tikz} 
\newcommand{\orcidicon}{\includegraphics[width=0.32cm]{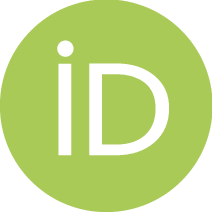}}
\foreach \x in {A, ..., Z}{%
\expandafter\xdef\csname orcid\x\endcsname{\noexpand\href{https://orcid.org/\csname orcidauthor\x\endcsname}{\noexpand\orcidicon}}
}

\begin{document}

\title{Cibersapiens: Gamifying Cybersecurity Education to Engage Secondary Students}

\author{\IEEEauthorblockN{
\orcidA{}Soraya Fr\'ias-Guadalupe$^{1}$,
\orcidB{}Pedro Paredes$^{2}$,
\orcidC{}Carlos Castaneda-Marroqu\'in$^{3}$
}

\IEEEauthorblockA{$^1$Facultad de Ciencias de la Educaci\'on y del Deporte y Estudios Interdisciplinares\\
s.friasg@alumnos.urjc.es}
\IEEEauthorblockA{$^2$Escuela T\'ecnica Superior de Ingenier\'ia Inform\'atica, Universidad Rey Juan Carlos\\
pedro.paredes@urjc.es}
\IEEEauthorblockA{$^3$Serval Networks, Av. de la Industria, 4, Alcobendas, 28108, Madrid, Spain\\
ccastaneda@servalnetworks.com}
}


\maketitle

\begin{abstract}
Cybersecurity education and awareness are often tailored for businesses, leaving general citizens, especially young people, unprepared. Despite the availability of educational resources, cybersecurity training is not provided equally to all individuals, particularly children and teenagers. As digital risks such as sexting, grooming, and cyberbullying, among others, increase, young users remain vulnerable. However, existing awareness efforts fail to engage them effectively and current content lacks the motivation needed to drive meaningful behavioral change.
This paper introduces "Cibersapiens: Digital Well-being for a Connected Generation," an interactive, gamified environment designed to teach cybersecurity to secondary school students. The project focuses on developing and implementing gamified educational resources that actively engage students, improve their understanding of cyber risks, and promote best practices in digital security. By leveraging active learning methodologies, this approach transforms cybersecurity education into an engaging and meaningful experience. 

\end{abstract}


{\bf Contribution type:}  {\it  Training and educational innovation}

\section{Introduction}\label{intro}
The increasing digitalisation of daily life has exposed young people to various cybersecurity threats. Despite existing educational resources, cybersecurity awareness among adolescents remains low, as traditional approaches do not engage them effectively \cite{rahman2020importance}. Many young users lack the necessary digital literacy to recognize and respond to online risks, making them particularly vulnerable to manipulation and exploitation \cite{gomez2020seguridad}. In addition, cybersecurity education is often fragmented and not adapted to the specific needs of teens, leading to a gap between knowledge and practical application. Addressing this issue requires innovative student-centred methods that not only provide theoretical knowledge but also foster behavioral change through engaging and interactive learning experiences \cite{prats2018diseno}.
Teenagers aged 12 to 16 are highly active online, yet they often underestimate digital risks \cite{martin2022percepcion}. Many do not recognize the consequences of unsafe behavior, making them vulnerable to cyber threats such as identity theft, misinformation, and online harassment \cite{al2019effectiveness}. Additionally, their digital interactions increasingly involve social networks, gaming platforms, and online communities, where personal information is frequently shared without appropriate safeguards \cite{jimenez2011perspectiva}. Although schools introduce basic digital literacy, these efforts are often insufficient, lacking engaging methodologies that resonate with students. As cyber threats evolve, there is a growing need for cybersecurity education that is interactive, relevant, and capable of fostering long-term behavioral change \cite{bada2019cyber}.


Gamification is an educational approach that incorporates game mechanics to enhance learning experiences \cite{chou2019actionable}. Research has shown that it increases motivation, engagement, and retention, making it a valuable tool in education \cite{roa2021evaluacion}. Applied to cybersecurity training, gamification can create immersive learning environments that help students develop awareness and digital security skills while making the learning process more enjoyable and interactive. Traditional cybersecurity education often relies on passive methods, such as lectures and informational materials, which do not capture the students’ attention or encourage long-term behavioral change \cite{williams2024leveraging}.


For secondary school students, particularly those aged 12 to 16, gamification is especially effective as it aligns with their natural inclination toward interactive and competitive activities. By embedding cybersecurity concepts into engaging challenges, role-playing exercises, and simulated cyber incidents, students develop a deeper understanding of online risks, such as phishing, social engineering, and privacy breaches. Ultimately, gamification transforms cybersecurity education into a meaningful and impactful experience that prepares students to navigate the digital world responsibly and securely.


This paper presents Cibersapiens: Digital Well-being for a Connected Generation, a gamified educational framework designed to teach cybersecurity to secondary school students. The project aims to address the significant gap in cybersecurity education for young people by integrating interactive, game-based learning into their daily educational experience. 


Through Cibersapiens, our aim is to transform the way cybersecurity is taught by moving away from passive, lecture-based instruction to active, hands-on learning. This transition is crucial because young people are increasingly exposed to digital dangers and traditional educational tools often do not prepare them adequately. By incorporating gamification, Cibersapiens not only enhances students' technical knowledge but also instils lasting behavioral changes that promote responsible and secure digital practices. Ultimately, this paper highlights the transformative potential of gamification in fostering cybersecurity awareness and skills among the next generation of Internet users.

The main contributions of this paper are summarized as follows:

\begin{itemize}
    \item A discussion on best practices for integrating gamification into cybersecurity education: This paper identifies and discusses effective strategies for integrating gamification into the curriculum. Drawing on lessons from Cibersapiens, we provide guidelines and recommendations on how to design gamified cybersecurity programmes that resonate with students, increase their engagement, and address specific challenges they face in the digital world. By outlining these best practices, the paper aims to support educators in adopting gamification as a powerful tool for teaching cybersecurity.
    \item An alignment of gamified cybersecurity education with national and European competence frameworks: This paper presents a structured review of how gamified cybersecurity education can align with the digital competence goals defined by the Spanish LOMLOE education law and the European DigComp framework. This ensures that the proposed activities are not only engaging, but also pedagogically sound and consistent with educational standards. By connecting learning outcomes with specific competence areas, the paper supports the integration of cybersecurity education into broader objectives like digital literacy, responsible use of technology, and the development of critical thinking.
    \item Design and implementation of Cibersapiens, a gamified cybersecurity learning environment: This paper presents the development and execution of Cibersapiens, an innovative and interactive platform that uses gamification to teach cybersecurity to secondary school students. The design focusses on creating an engaging and immersive learning environment where students can actively participate in realistic cybersecurity challenges and scenarios, helping them better understand digital risks and best practices for online security. In adittion, the implementation of Cibersapiens takes into account the participation of the entire educational community, including teachers, parents, and school administrators, ensuring a holistic approach to fostering digital well-being and cybersecurity awareness.


\end{itemize}

This paper is structured as follows. Section ~\ref{motivation} provides an overview of effective ways to increase awareness of digital risks and emphasizes the importance of addressing these issues early in students' lives. ~\ref{sota} reviews existing literature on cybersecurity awareness in secondary education, highlighting the challenges and current approaches to teaching digital security. \ref{method}, introduces the design and implementation of Cibersapiens, detailing the educational principles and gamification techniques applied on the platform. Section \ref{Insights}, presents initial feedback from educators and students, offering valuable insight into the effectiveness and engagement of the gamified approach. Finally, Section \ref{conclusions}, Conclusions and Future Work, summarizes the findings of the study and outlines directions for future research, focussing on refining the platform and expanding its application to other educational contexts.

\section{Motivation}\label{motivation}





The teaching of cybersecurity to adolescents aged 12 to 16 is motivated by several crucial factors. At this age, young people are highly connected to the digital world. Understanding cybersecurity empowers them to protect their personal information, avoid identity theft, and mitigate risks such as cyberbullying or online fraud \cite{livingstone2017maximizing}. In addition, training young people in this area promotes responsible digital citizenship by fostering an understanding of the impact of their online actions and promoting respect for the privacy and rights of others \cite{ribble2011digital}.

Educating adolescents about cybersecurity also indirectly contributes to a more secure digital ecosystem. Reducing vulnerable behaviours minimises the likelihood of cyberattacks on a community or even global scale \cite{bada2019cyber}. Moreover, digital literacy is essential for future careers, not only in technology-related fields, but in all sectors of an increasingly digital society \cite{OECD2019skills}. Cybersecurity education encourages risk analysis, critical thinking, evaluation of information sources, and informed decision-making skills that extend far beyond the digital realm.

Despite these benefits, the exponential growth in the use of digital technologies by adolescents has not been accompanied by a parallel increase in cybersecurity awareness or digital responsibility. Although cybersecurity-related content has been formally included in the curriculum of the Community of Madrid, studies reveal that adolescents remain highly vulnerable to online risks such as cyberbullying, identity theft, sextortion, and misinformation. Traditional educational approaches, often based on theoretical sessions or isolated workshops, do not engage students effectively or promote long-term behavioural change.

Recent literature highlights the scarcity of comprehensive evidence-based cybersecurity education programs. Among the few that exist, even fewer implement active learning methodologies or gamification, despite the documented benefits of these strategies in terms of student motivation, meaningful learning, and knowledge retention. Additionally, the lack of programs targeting both students and their families represents a critical gap, considering that adolescents often depend on adults who themselves lack sufficient training in safe digital practices.

This situation calls for the design and implementation of innovative, dynamic and empirically grounded educational interventions capable of reaching young learners in a language and format that resonate with their interests and lived experiences. Gamification, as an active pedagogical approach, offers a powerful means of turning cybersecurity education into an engaging and impactful experience. In this context, Cibersapiens emerges as a novel educational proposal that integrates digital citizenship, cybersecurity, and game-based learning to empower secondary school students to make safe, ethical, and responsible decisions in the digital world.

\section{State of the art}\label{sota}

\subsection{Gamification}
Gamification involves the application of game mechanics, elements, and design techniques in non-game contexts to engage users and address problems.\cite{zichermann2011gamification} \cite{werbach2012for}. This active methodology has experienced significant growth in recent years, placing the student at the center of the learning process in an autonomous and participatory manner. ~\cite{jara2022estrategias}. Active teaching methodologies are grounded in the constructivist pedagogical model, providing an appealing alternative to traditional education by placing greater emphasis on what the student learns rather than what the teacher instructs. This approach fosters deeper understanding, increased motivation, and greater participation of the student in the learning process.~\cite{taipe2020metodologias}

Several authors have identified the main advantages of this methodology, highlighting an increase in student motivation, the creation of a safe learning environment, increased student awareness of their progress, the promotion of cooperation, the improvement of self-awareness with respect to abilities and the enhancement of retention of acquired knowledge in a meaningful way.\cite{roa2021evaluacion}

Gamification emerges as one of the most effective methodologies for teaching cybersecurity content, outperforming traditional methods currently in use, which have shown limited impact among adolescents. In this regard, the study “Leveraging Gamification and Game-based Learning in Cybersecurity Education: Engaging and Inspiring Non-Cyber Students” highlights the suitability of gamification to engage and inspire students who are not specialized in cybersecurity. Practical experiences, such as "Capture the Flag" (CTF) events, are particularly effective as they enable students to solve problems in a dynamic and appealing environment. \cite{williams2024leveraging} 

\subsection{Cybersecurity Education in the Curriculum of the Community of Madrid}
In the Community of Madrid, the subject "Ciencias de la Computación", an elective in 1st and 2nd year of ESO, develops specific competency 4: understanding how computer systems communicate to form networks and the importance of cybersecurity.
\begin{itemize}
\item 1st ESO: Safe and responsible use of the Internet.
\item 2nd ESO: Cybersecurity (Internet security, types of attacks, threat identification, and protection software).
\end{itemize}

The subject "Tecnología y Digitalización", mandatory in 2nd and 3rd year of ESO, focuses on:
\begin{itemize}
\item Competency 6: Understanding how devices and applications work, optimizing their use, and solving technical issues.
\item Competency 7: Using technology responsibly, assessing its impact.
\end{itemize}
In 2nd ESO, the digitalization of the personal learning environment is covered:
\begin{itemize}
\item Safe Internet use: information search, email, messaging, and social networks.
\item Learning tools and platforms: configuration, maintenance, and critical use.
\item Network security: risks, threats, attacks, and protection measures (antivirus, firewalls, proxy).
\item Good digital practices: cyberbullying, sextortion, privacy, inappropriate content, and addiction risks.
\end{itemize}
In 4th ESO, the subject "Digitalización" (Technological option) includes:
\begin{itemize}
\item Competency 3: Digital well-being habits, device and data protection, and health.
\item Competency 4: Critical digital citizenship and responsible use of technology.
\end{itemize}
The following topics are covered:
\begin{itemize}
\item Digital security and well-being: Introduction to cybersecurity, device security (antivirus, backups), data protection (digital identity, privacy, GDPR), and health risks (ergonomics, mental well-being).
\item Critical digital citizenship: Freedom of expression and its limits, netiquette, intellectual property, digital journalism, fake news detection, online administrative procedures (electronic ID, digital signature), secure e-commerce, and ethics in data usage (AI, planned obsolescence, digital sustainability).
\end{itemize}

\subsection{Cybersecurity Education Programs}
Training in digital skills has not kept pace with the rapid technological development, leading to significant gaps in adolescents' knowledge about how to protect themselves in digital environments. Despite the increasing use of technology, education about its safe and responsible use has not progressed at the same rate. As a result, the responsibilities associated with the intensive use of ICT are not adequately addressed \cite{yellowlees2007problematic}.

Despite the wide availability of cybersecurity content in high-quality repositories, such as those offered by INCIBE, the real gap lies in the lack of effective educational proposals that use methodologies capable of making a real impact on students and helping them identify digital risks. One-off initiatives, such as workshops or informational talks, often lack the appeal and effectiveness of more dynamic and interactive approaches, such as gamification and game-based learning.

To assess the current state of cybersecurity education, searches were performed in academic databases such as Dialnet, Scopus, and ERIC, as well as in search engines such as Google Scholar. The goal was to identify studies that focus on preventive actions implemented in educational settings. However, the number of relevant results was limited and increased further when filters related to innovative methodologies were applied.

This shortage of programs promoting responsible Internet use and digital risk management is reflected in the study "Responsible Internet Use and Digital Security: A Systematic Review of Educational Programs". This research highlights the limited number of published empirical studies that examine existing educational programs in this area and evaluate their results. Of a total of 702 articles initially considered, only 19 were selected for relevance. Although international organizations have proposed initiatives to address digital risks, very few educational programs have been developed on the topic. Furthermore, the scientific literature stresses the urgent need to design and implement training programs on digital security and responsible technology use aimed at the entire educational community \cite{aznar2024uso}.


The article highlights that cybersecurity education programs have been implemented primarily in Spain (36.8\%) and the United States (15.8\%). Of these, 68.4\% targeted students, mainly children and young people up to 13 years of age. Most of the programs were sort, usually lasting a few weeks, with only a few designed as continuous interventions. Common topics addressed included cyberbullying, sexting, grooming, identity theft, and misinformation.

Among the 19 programs analyzed, only one incorporated gamification as a strategy to evaluate learning about privacy concepts. It is particularly concerning that none of these programs were aimed at families, despite the fact that children often rely on adults for guidance on safe Internet use, although many of these adults lack the necessary training themselves \cite{aznar2024uso}.

In this regard, Prats Fern\'andez et al.\cite{prats2018diseno} emphasize that, although there have been efforts to develop programs aimed at promoting responsible use of ICT, their implementation in real educational contexts remains limited. This has resulted in a scarcity of publications evaluating the effectiveness of preventive measures and interventions related to ICT-related risks. The authors stress the importance of continuing to promote studies that provide relevant data on the development and impact of such measures for adolescents. Furthermore, they highlight a clear need and social demand for ongoing education in this area to raise awareness and reduce the risks associated with technology.


It is essential to develop meaningful work in cyber education to help preadolescents understand the risks associated with the digital environment. The creation of interactive and gamified environments emerges as one of the most effective strategies to teach cybersecurity, with a purpose that goes beyond simple awareness. The goal is to design training programs that encourage children to adopt good Internet security practices, transforming them into habits through a process of assimilation that integrates technological actions into their daily behavior through continuous practice.

\section{Method}\label{method}
\subsection{Educational intervention}

\textit{Cibersapiens: Digital Wellbeing for a Connected Generation},  is an interactive and gamified environment aimed at enhancing cybersecurity education at the lower secondary level. This initiative focuses on the design and implementation of gamified instructional resources that foster student engagement and a deeper understanding of cyber risks and responsible digital practices, through the application of active learning methodologies. The overarching goal is to transform cybersecurity instruction into an engaging and meaningful educational experience that motivates students to take an active role in their learning process, while fostering greater awareness and long-term involvement in the entire educational community.

\subsection{Contents}

 The content of the proposal is structured around three fundamental pillars: conceptual, procedural, and attitudinal. The first module, "Introduction to Cybersecurity," introduces key cybersecurity concepts and highlights the importance of protecting personal information in digital environments. The second module, "Risks in the Digital Environment," focusses on the procedures needed to identify and manage the main risks faced by adolescents, including cyberbullying, grooming, sexting, loss of privacy, and security threats such as malware, phishing, and ransomware. This module also addresses the impact of disinformation, how to detect fake news, and the symptoms associated with technology addiction, particularly its consequences for mental and physical health, along with strategies for achieving a healthy and balanced use of technology.
Finally, the third module, "Prevention and Protection Strategies," emphasizes the importance of fostering responsible and ethical attitudes in the use of technology, encouraging critical and reflective thinking in the face of digital challenges. Covers the responsible use of social networks, privacy settings, how to handle risky situations (including support resources and who to turn to), and offers practical tools to begin learning about cybersecurity.

\begin{table}[H]
\centering
\begin{tabular}{|p{1cm}|p{2.7cm}|p{2.7cm}|}
\hline
\textbf{Session} & \textbf{Thematic Block} & \textbf{Content} \\
\hline
1 & Introduction to cybersecurity & Key concepts and their importance \\
\hline
2 & Risks in the digital environment & Cyberbullying, grooming, and online harassment \\
\hline
3 & Risks in the digital environment & Sexting and loss of privacy \\
\hline
4 & Risks in the digital environment & Security threats (malware, phishing, ransomware) and disinformation \\
\hline
5 & Risks in the digital environment & Technology addiction and its impact on physical and mental health \\
\hline
6 & Prevention and protection strategies & Cybersecurity education and privacy settings \\
\hline
7 & Prevention and protection strategies & Responsible use of social networks and managing risky situations \\
\hline
8 & Wrap-up and assessment & Final reflection and learning evaluation \\
\hline
\end{tabular}
\caption{Session plan for the cybersecurity education program}
\label{table1}
\end{table}

 As you can see in table \ref{table1}, the content of the program is organized into eight 50-minute sessions, aligned with the standard school schedule. These sessions are structured around three thematic blocks: (1) Introduction to cybersecurity, (2) Risks in the digital environment, and (3) Prevention and protection strategies. This organization enables a progressive exploration of the conceptual, procedural, and attitudinal dimensions of cybersecurity education. The detailed session plan is described below:


Through the proposed content, the program promotes the development of key competencies aligned with the Secondary Education curriculum (ESO), following Decree 65/2022 of July 20, issued by the Governing Council, which establishes the organization and curriculum for this educational stage in the Community of Madrid. In particular, this intervention enables students to strengthen essential competencies such as digital competence, learning to learn, citizenship (social and civic) competence, and linguistic communication skills. These capabilities are crucial for students to communicate effectively in digital environments and to become active and responsible citizens in today’s society.



This content proposal is also closely aligned with the Digital Competence Framework for Citizens (DIGCOMP), an essential European Union-level tool for fostering digital competence among citizens. Specifically, it focusses on the areas of security (area 4) and problem solving (area 5), aiming to equip students with the ability to protect their devices, content, personal data, and privacy in online environments. Furthermore, it seeks to raise awareness of physical and mental health about the use of technology and provide the skills necessary to identify needs and solve problems in digital contexts.


 The primary goal is for students to develop these competencies by progressing from a basic level with teacher guidance to a basic level with autonomy, while still receiving support when needed. The initiative aims to provide students with educational materials that promote an initial level of cybersecurity awareness. Although valuable repositories such as the one offered by the Spanish National Cybersecurity Institute (INCIBE) can be used as complementary resources, this intervention is distinguished by its emphasis on a formative rather than merely informative approach. Through the combination of educational materials and active methodologies such as gamification, the initiative seeks to generate a meaningful impact on students, promoting an initial level of awareness of the responsible and safe use of technology.
 

 In this regard, a recurring educational design is proposed for schools, one that involves not only students but also the broader educational community, with particular emphasis on families and teaching staff. This comprehensive approach aims to foster a culture of cybersecurity awareness and best practices that go beyond the classroom and reach the social environment of students.
 

With regard to teachers, the contents of this intervention proposal are designed to support their training in digital competencies. These materials are expected to help teachers reach the A2 proficiency level, as well as raise their awareness of the importance of these skills, enabling them to effectively transfer this knowledge to their students during their lessons.
 

Likewise, the training proposal extends its reach to students’ families, recognizing their essential role in the development of digital competencies. The content will provide families with clear information and practical tools to understand the risks and opportunities inherent in the digital world. In this way, the objective is to empower parents and guardians to actively support their children to acquire safe, responsible, and critical habits in the use of technology, fostering a family environment that promotes positive digital citizenship.

In the CiberSapiens: Digital Wellbeing for a Connected Generation proposal, gamification stands out as the main methodological strategy for teaching cybersecurity due to its ability to increase student motivation and engagement. By incorporating game elements into the teaching-learning process, students are expected to feel more interested and willing to participate in cybersecurity-related activities. Furthermore, gamification allows complex topics to be addressed in a playful way, facilitating both understanding and information retention.

This proposal also promotes active learning and student engagement through practical activities, debates, and collaborative projects. Encourage teamwork, critical thinking, and cooperation among students to solve problems and share knowledge, fostering cooperative learning.

\subsection{Materials and Platforms}

This intervention proposal relies on a variety of resources aimed at maximizing adolescent engagement and understanding in the field of digital security.

First, it is essential to have adequate technological equipment, such as desktop computers, laptops, and tablets, that allows students to access content and actively participate in the proposed activities. In addition, it is crucial to ensure high-speed internet connectivity in classrooms and/or shared spaces where sessions take place, thus providing a seamless and uninterrupted experience. It is also important to have a projector and screen for theoretical presentations, as well as classrooms equipped with appropriate furniture to support both group and individual activities.

To complement face-to-face activities, the use of a virtual learning platform is proposed, such as EducaMadrid. This platform would function as a repository for materials, a discussion forum, and a tool to monitor student progress. It allows students to access content anytime and anywhere, thus promoting autonomous learning and peer collaboration.

With regard to teaching materials, the creation of attractive and interactive resources is supported by tools such as Canva and Genially. These resources include presentations, infographics, videos, and gamified activities that allow students to learn in an engaging and participatory way. It is also recommended that you have stationery and office supplies available for the creation of informational posters and other supporting materials.

Gamification plays a central role in this proposal, with tools like Educaplay suggested to design engaging activities and interactive assessments that motivate students to learn and apply cybersecurity concepts. These tools allow the creation of games, quizzes and challenges customized to the level of competence of the students, thus encouraging personalized learning.

Moreover, communication and collaboration tools such as Wikis and Padlet are recommended to facilitate interaction among members of the educational community and collaborative work on cybersecurity-related projects. These tools encourage students to share ideas, resources, and experiences, contributing to a collective understanding of digital security.

To support the development of sessions, specific games have been designed to encourage active participation and practical learning (see table \ref{table2}). These games, which are described below, are integral to the proposal and essential to achieving awareness and learning goals. Cybersecurity Word Roulette, Crossword Puzzle, Role-Playing Game or Escape Room, Digital Detectives, Ciberquiz: Defend Your Digital Space, Privacy Puzzle: Sequencing Strategies, Interactive Map: Exploring Threats and Misinformation, Froggy Challenge on Tech Addiction, Practical Workshop: Configure Your Social Media for Safe Browsing, and Matching Game: Safety in the Digital World.

Lastly, access to Open Educational Resources (OER), such as those available on Procomún and other repositories, as well as the resources provided by INCIBE, is recommended. These materials offer high-quality curriculum-aligned content suitable for various educational levels and contexts, allowing teachers to enrich their sessions and students to deepen their knowledge in areas of personal interest.


\begin{table}[h]
    \centering
    \renewcommand{\arraystretch}{1.2} 
    \begin{tabular}{|p{3.9cm}|p{3.9cm}|}
        \hline
        \textbf{Session} & \textbf{Game} \\ 
        \hline
        1. Introduction to cybersecurity & a. Word Roulette (pre-test) \\ 
        & b. Crossword (pre-test) \\ 
        \hline
        2. Risks in the digital environment & a. Escape Room \\ 
        & b. Cyberquiz \\
        \hline
        3. Sexting and loss of privacy & a. Sequence ordering game \\ 
        \hline
        4. Security threats and \\Disinformation & a. Interactive map\\ 
        \hline
        5. Technology addiction & a. Froggy Challenge\\ 
        \hline
        6. Cybersecurity education and privacy settings & a. Practical workshop\\ 
        \hline
        7. Responsible use of social media & a. Matching situations \\ 
        \hline
        8. Closing and evaluation & a. Word Roulette (post-test) \\ 
        & b. Crossword (post-test) \\ 
        \hline
    \end{tabular}
    \caption{Session and Game Table}
    \label{table2}
\end{table}



\subsection{Sample Activity}

Below is a sample gamified activity designed for Session 2, “Cibersapiens in Action: Defend Your Digital Space from Grooming and Cyberbullying!” This session aims to address the risks associated with cyberbullying and grooming, raise student awareness, and promote active learning and critical reflection.

The session begins with a brief introduction that outlines its objectives, followed by a theoretical presentation. In this section, the concepts of cyberbullying, grooming, and online harassment are introduced. The various ways in which these phenomena manifest, their consequences, and prevention strategies are discussed.

Next, the practical part of the session begins, focussing on the game “Digital Detectives.” This activity allows students to explore the topics in a playful and participatory manner, promoting empathy and responsibility in the digital environment.

In “Digital Detectives,” students take on the role of rookie detectives with the mission of rising through the ranks to become professional digital agents. Using the Genially platform, students can access an interactive map that presents four critical missions. In each mission, they face questions and scenarios that require identifying risks, making informed decisions, and applying prevention strategies. Upon completing each mission, students receive clues that allow them to progress and, ultimately, unlock a virtual briefcase by cracking a secret combination.



The first students to solve the puzzle are rewarded with the "Cybersecurity Kit," which includes educational resources, practical guides, and a “Professional Digital Agent” diploma — a symbolic recognition celebrating their achievement in the activity. This dynamic not only encourages learning about the dangers of cyberspace but also promotes critical thinking and teamwork.

The main goal of this game is to help students identify and better understand the dynamics of cyberbullying and grooming, as well as to develop skills to make informed and responsible decisions when facing these situations. Through this role-playing game, students gain practical insight into how these threats operate in the digital environment, what warning signs to look out for, and what actions they can take to protect themselves and others.

The activity fosters empathy by placing students in the position of victims and encouraging reflection on the consequences of their actions in the digital world. By assuming the role of detectives, students must analyze complex scenarios, assess risks, and make decisions that affect the well-being of others, helping them develop greater awareness of the importance of acting responsibly and compassionately in cyberspace.




After the “Digital Detectives” activity, the students participate in a second game: “Cyberquiz: Defend Your Digital Space.” In this activity, students access an interactive online quiz featuring definitions and questions related to grooming, cyberbullying, and online harassment. Working in teams, they participate in a competitive quiz designed to assess their understanding of the concepts covered. In the end, the answers are reviewed, and key definitions and ideas are discussed, including topics such as grooming, cyberbullying, sexting, and data privacy, which encourage teamwork and collaboration. The goal of this activity is to reinforce the students' understanding of the key terms and concepts related to online abuse, grooming, and cyberbullying.

Finally, the session concludes with a group reflection in which the students share their thoughts on cyberbullying and discuss possible intervention strategies, thus consolidating what they have learnt during the session. In conclusion, both “Digital Detectives” and “Cyberquiz” serve as playful and effective tools to raise awareness among students about the dangers of cyberbullying and grooming, provide them with practical strategies to prevent and address these issues, and promote a culture of respect and responsibility in the digital world.

\subsection{Engaging the Educational Community}

Cybersecurity is a crucial aspect of adolescent education, and its resolution requires active collaboration of the entire educational community, including families and teachers. This intervention proposal aims to create a learning environment in which parents, educators, and experts work together to protect young people in the digital world. Families and teachers are key figures in adolescents' daily lives, and their influence can be decisive in shaping healthy digital habits. Therefore, it both groups must have access to tools and resources related to cybersecurity. Teachers can provide effective training, while families can encourage responsible use of technology at home.



To support this collaboration, a blog titled “Cibersapiens: The Cybersecurity Blog for Parents and Educators” has been created. This platform aims to provide clear, concise, and practical information on online risks and strategies to prevent them. The blog empowers families and teachers with the knowledge they need to guide teens in the safe and responsible use of technology. It is intended for parents, teachers, and educators who care about the well-being of adolescents in the digital environment. The blog includes content designed for students, as well as additional materials, such as infographics, videos, and relevant news.

Among the resources offered are articles on the most common risks in the digital environment, practical guides to configure privacy on social networks and apps, tips for encouraging a balanced use of technology, educational resources and tools to dive deeper into cybersecurity, best practices and examples of how other parents and teachers are addressing cybersecurity challenges, a discussion forum to share experiences and advice, information about events, workshops, and courses, and contact information for cybersecurity experts and support resources.



In addition to the blog, several activities are proposed within the school to engage the entire educational community. These include awareness campaigns, using posters, leaflets, and videos to raise awareness of online risks and encourage responsible technology use; events such as talks, workshops, and exhibitions aimed at students, families, and teachers, delivered by cybersecurity experts; and the creation of a cybersecurity policy, establishing clear rules and procedures to protect the data and privacy of all members of the educational community.

Active participation of families and teachers in cybersecurity education is essential to build a safe and responsible digital environment. Through collaboration and access to appropriate resources, adolescents can be empowered to navigate the digital world in a safe, ethical, and mindful manner.

\subsection{Evaluation Design}
The evaluation of the intervention will be structured into three phases: initial, procedural, and final, using a combination of quantitative and qualitative methods. Conceptual learning will be assessed through statistical analysis of data from the Genially and Educaplay platforms, based on students' performance in gamified activities specifically designed. During the initial phase, a pre-test will be administered using a roulette-style game and a crossword puzzle focused on core cybersecurity concepts. These same activities will be repeated in the final phase as a post-test, enabling comparison and measurement of knowledge gains.

Practical and attitudinal learning will be evaluated through direct observation, supported by a structured checklist and a learning diary maintained by the students, in which they will document their experiences, reflections, and observed behaviors throughout the intervention. Finally, a satisfaction questionnaire will be administered to students, along with a separate survey to teachers, aimed at collecting feedback on the perceived effectiveness of the intervention and identifying areas for improvement. 
\section{Insights and Preliminary Findings}\label{Insights}

The preliminary implementation of Cibersapiens, a gamified cybersecurity learning environment designed for secondary education, has provided valuable insight into both the current state of digital security education and the potential for innovative pedagogical interventions. Although full deployment and data collection phases are ongoing, early findings provide a strong foundation for reflection, refinement, and future scalability.

A key insight concerns the limited and fragmented presence of cybersecurity topics within the official curriculum. Digital security is often treated as a peripheral subject, typically addressed within optional or highly specialized courses such as Technology or Computer Science, and not systematically integrated into core subjects. In many schools, these topics appear only in the second or third year of compulsory secondary education (ESO), if at all. This curricular gap creates a scenario in which many students complete their formal education with minimal exposure to essential knowledge and practices related to cybersecurity.


Moreover, a review of existing literature and initial field observations highlights a recurring shortcoming in previous initiatives: the lack of continuity, contextualization, and pedagogical depth. Many programs rely heavily on isolated workshops or awareness days, which, while useful, are insufficient to instill sustained behavioral change or deep understanding. Cibersapiens addresses this challenge by proposing a more holistic and embedded model, tightly aligned with curricular standards, and tailored to the developmental needs of adolescents.


The proposed approach combines three interrelated components: (1) a gamified learning environment; (2) a collaborative blog for the educational community; and (3) a school-wide awareness campaign. The use of gamification—through role-playing scenarios, quizzes, and interactive challenges—has emerged as a particularly promising strategy. 
These tools engage students in active, contextual learning, encouraging ethical reflection and internalization of safe digital habits. Narrative design and interactivity have proven to be effective in increasing motivation and supporting knowledge retention.

The initial pilot of these gamified activities suggests that interactivity and narrative design significantly enhance student motivation. Role-playing exercises, for instance, have allowed students to explore multiple perspectives (e.g. hacker, victim, bystander), promoting empathy and a deeper understanding of the implications of digital behavior. Similarly, quizzes and minigames have been effective in consolidating technical concepts such as password hygiene, phishing detection, and privacy settings.These learning activities have been well-received by both students and teachers, according to initial qualitative feedback collected through observations and post-session discussions.

One of the most important insights so far is the need to broaden the focus of cybersecurity education beyond the student body. Families and educators play a fundamental role in reinforcing safe practices and supporting students' learning processes. The inclusion of parents and teachers in the initiative through a dedicated blog titled Cibersapiens: A Cybersecurity Blog for Parents and Educators has been designed to build a shared knowledge base and foster a culture of digital responsibility across the entire school community.

The blog acts as both a resource hub and a participatory platform. It provides access to articles, infographics, tutorials, videos, and links to external resources. It also encourages educators and parents to contribute, turning them from passive recipients of information into active agents of cybersecurity education. In its future development, the blog is envisioned as a wiki-style collaborative space, allowing co-creation of content and fostering stronger community ownership of digital safety initiatives. These tools are expected to support the continuous refinement of both content and delivery methods.
Nevertheless, sustaining the relevance of content in the face of emerging digital threats will require regular updates and institutional commitment.


Although the project is still in its early implementation stages, another key insight has emerged around the need for robust evaluation tools. To assess the impact of the intervention, it is essential to go beyond knowledge tests and incorporate multidimensional strategies capable of capturing behavioral change, self-efficacy, and perceived risk.

Future phases of the project will prioritize the development of comprehensive assessment instruments, including pre- and postintervention questionnaires, classroom observations, and analytics collected through the gamified platform (e.g., time on task, engagement patterns, decision-making paths). These tools are expected to support the continuous refinement of both content and delivery methods.

Furthermore, feedback loops involving students and teachers are being established to ensure that materials are adaptable to different learning contexts and that activities are perceived as meaningful and culturally relevant.

Another crucial insight relates to the pivotal role of teachers in the success of cybersecurity education initiatives. Although Cibersapiens includes materials and activities aimed at the school community, preliminary observations suggest that many educators feel underprepared to address cybersecurity topics with confidence. This is in part due to the lack of formal training in digital safety and the absence of structured professional development opportunities in this area. As such, empowering teachers through targeted training, ongoing support, and collaborative resource design is essential. Their active involvement not only ensures the continuity and contextualization of learning experiences but also helps foster a school culture where cybersecurity becomes a shared responsibility.

In summary, the preliminary findings of the Cibersapiens initiative underscore the importance of an integrated, participatory, and pedagogically grounded approach to cybersecurity education. The combination of gamified learning, community engagement, and continuous feedback mechanisms appears to offer a promising path forward. These early results will guide the refinement of the intervention and lay the groundwork for its broader application in secondary schools. 
\section{Conclusions and Future Work}\label{conclusions}

The development of Cibersapiens highlights the potential of gamified environments as a meaningful response to the growing need for cybersecurity education in secondary schools. By aligning with active learning methodologies and involving the entire educational community, the project fosters both student engagement and collective digital awareness.

Preliminary observations suggest that integrating cybersecurity into the curriculum through experiential and interactive formats can effectively promote critical reflection, responsible online behavior, and long-term attitude change. Furthermore, the project underscores the importance of inclusive strategies that involve not only students but also educators and families.

Future work will focus on refining the proposed activities, expanding the blog into a collaborative platform, and implementing robust evaluation tools to assess learning outcomes and behavioral impact. These steps will contribute to shaping a scalable and adaptable model for cybersecurity education in schools.

\section*{Acknowledgment}
This work is part of the project I+D+i PID2022-137849OB-I00 and has been supported by MICIU/AEI/10.13039/501100011033/ y FEDER/UE

\bibliographystyle{IEEEtran}
\bibliography{bibliography}

\end{document}